\documentclass[%
  journal=jctcce,%
  manuscript=article,%
  layout=twocolumn,%
]{achemso}
\setkeys{acs}{maxauthors=3, etalmode=truncate}
\usepackage[T1]{fontenc} 
\usepackage{graphicx}
\usepackage{amssymb}
\usepackage{booktabs}
\usepackage{array}
\usepackage{mathtools}
\usepackage{amsmath}
\usepackage{upgreek}
\usepackage{float}
\usepackage{color}
\usepackage[range-phrase=\text{--}, range-units=single]{siunitx}
\DeclareSIUnit\angstrom{\text{Å}}
\usepackage{comment}

\usepackage{caption}
\usepackage{ifthen}
\makeatletter
\newcommand{\isSingleColumn}{\equal{\acs@layout}{traditional}}
\newcommand{\isSuppinfo}{\equal{\acs@manuscript}{suppinfo}}

\ifthenelse{\isSingleColumn}%
    {
        \usepackage{setspace}
        \AtBeginDocument{\singlespacing}

        \renewcommand{\maketitle}{\section*{\@title}}
        \usepackage[fontsize=11pt]{fontsize}
    }{
	    \usepackage[fontsize=9pt]{fontsize}
    }%

\ifthenelse{\isSuppinfo}%
  {%
  	\setlength{\bibsep}{4pt plus 0.3ex}
  	\DeclareCaptionTextFormat{baseline}{\baselineskip 4.5mm #1}
  }{%
    \setlength{\bibsep}{0pt plus 0.3ex}%
    
    \DeclareCaptionTextFormat{baseline}{\baselineskip 9.5pt #1}
  }%

\makeatother

\newcommand{\onlinecite}[1]{\hspace{-1 ex} \nocite{#1}\citenum{#1}} 

\graphicspath{{Figures/}{SIFigures/}}
\renewcommand{\vec}[1]{{\boldsymbol{#1}}}

\newcommand{\dir}{../Bib}

\newcommand{\SItabmemory}{S1}

\newcommand{\SIfigITScontrol}{S1}
\newcommand{\SIfigVillinMemory}{S2}
\newcommand{\SIfigShortTrajmodel}{S3}

\author{Sofia Sartore}
\author{Franziska Teichmann}
\author{Gerhard Stock}
\email{stock@physik.uni-freiburg.de}
\affiliation{Biomolecular Dynamics, Institute of Physics,
   University of Freiburg, 79104 Freiburg, Germany}
\date{\today}
\title{Markov-type state models to describe non-Markovian dynamics}

\begin{document}

\begin{abstract}

  When clustering molecular dynamics (MD) trajectories into a few
  metastable conformational states, the assumption of timescale
  separation between fast intrastate fluctuations and rarely occurring
  interstate transitions is often not valid. Hence, when we construct
  a Markov state model (MSM) from these states, the naive estimation
  of the macrostate transition matrix via simply counting transitions
  between the states may lead to significantly too short implied
  timescales and thus to too fast population decays. In this work, we
  discuss advanced approaches to estimate the transition
  matrix. Assuming that Markovianity is at least given at the
  microstate level, we consider the Laplace-transform based method by
  Hummer and Szabo, as well as a direct microstate-to-macrostate
  projection, which by design yields correct macrostate population
  dynamics. Alternatively, we study the recently proposed quasi-MSM
  ansatz of Huang and coworkers to solve a generalized master
  equation, as well as a hybrid method that employs MD at short times
  and MSM at long times. Adopting a one-dimensional toy model and an
  all-atom folding trajectory of HP35, we discuss the virtues and
  shortcomings of the various approaches.

\end{abstract}

%
%
\section{Introduction}

Classical molecular dynamics (MD) simulations are a versatile tool for
gaining insights into complex biomolecular
processes.\cite{Berendsen07} To facilitate the interpretation of the
ever-increasing amount of data obtained from such simulations, coarse
grained models such as Langevin equations\cite{Lange06b, Hegger09,
  Ayaz21} and Markov state models \cite{Buchete08, Bowman09,
  Prinz11,Bowman13a} (MSMs) can be useful. Interpreting MD
trajectories in terms of memoryless transitions between metastable
conformational states, MSMs are popular because they provide a
generally accepted state-of-the-art analysis, promise
to predict long-time dynamics from short trajectories, and are
straightforward to build using open-source
packages. \cite{Scherer15,MSMBuilder,Nagel23a}
The usual workflow to construct an MSM
consists of (i) selection of suitable input coordinates or features,
(ii) dimensionality reduction from the high-dimensional feature space
to some low-dimensional space of collective variables, (iii)
geometrical clustering of these low-dimensional data into microstates,
(iv) dynamical clustering of the microstates into metastable
conformational states (or macrostates), and (v) estimation of the
transition matrix associated with these states.

Here, we are concerned with the last step of the procedure, that is,
we want to discuss various approaches (and propose new ways) to
estimate the macrostate transition matrix $\vec{T}(t)$.  Since these
formulations may go beyond the scope of a common MSM, we call them
'Markov-type state models.'  In what follows, we assume that steps (i)
- (iv) resulted in a partitioning of $N$ macrostates, such that the MD
data can be converted to a state trajectory $I(t)$.  Moreover we
assume that all states are connected such that the MSM is ergodic, and
that the MD data represent a stationary and time-homogeneous
process. To focus on the effects of non-Markovianity, we for now also
assume sufficient sampling of the MD data (but see Sec.\ 3.2).

The macrostate transition matrix $\vec{T}(t)$ contains the
probabilities $T_{IJ}$ that the system jumps from state $J$ to state
$I$ within some pre-chosen lag time, which can be obtained by simply
counting the transitions from $J$ to $I$ within $t$.  When we denote
the time-dependent state vector by $\vec{P}(t)= (P_1,\ldots,P_N)^T$
with state probabilities $P_I$, the time evolution of the state model
can be written as
\begin{equation}\label{eq:Pt}
\vec{P}(t)=\vec{T}(t)\vec{P}(0).
\end{equation}
We note that the time-dependent transition matrix $\vec{T}(t)$ does
not involve a dynamical approximation (such as the Markov
approximation discussed below), but is directly obtained from the MD
data. As a consequence,
its practical use is limited by the fact that
the calculation of $\vec{T}(t)$ is restricted to times
$t \lesssim t_{\rm max}$, where $t_{\rm max}$ is the length of the MD
trajectories. That is, Eq.\ (\ref{eq:Pt}) is in essence a state
representation of the dynamics, but it does not allow any prediction
beyond the timescale of the MD simulations.

The latter can be achieved by invoking a Markov approximation, which
assumes a timescale separation between fast intrastate fluctuations
and rarely occurring interstate transitions. Assuming that the
intrastate fluctuations randomize for times longer than some specific
lag time $\tau$, we expect that the transition matrix $\vec{T}(t)$
becomes constant for $t \gtrsim \tau$. Hence, $\vec{T}(t)$ can be
approximated by
\begin{equation}\label{eq:MSM}
\vec{T}(t\!=\!m\tau) \approx \vec{T}^m(\tau)
\end{equation}
with $m=1,\,2 \ldots$, which only requires short MD trajectories (in
principle as short as $\tau$) to estimate the transition matrix.
Since the time evolution of the MSM is given in steps of $\tau$, the
lag time defines the time resolution of the model, and therefore needs
to be chosen shorter than the fastest dynamics of interest. This
condition, however, is often in conflict with the above requirement of
a long enough lag time to achieve Markovianity of the system, i.e., a
constant transition matrix. Therefore, in practice one often ends up
with a state partitioning that is only approximately Markovian at the
desired time resolution, which raises the question of the optimal MSM
that best approximates the dynamics in this case.

There are various ways to compute the MSM transition matrix
$\vec{T}(\tau)$, in order to optimize the approximation of the true
transition matrix $\vec{T}(t)$ in Eq.\ (\ref{eq:MSM}). For one, we may
go beyond the simple counting scheme mentioned above and invoke our
knowledge of the microstates the macrostates are built
of.\cite{Roeblitz13,Hummer15,Kells19,Sharpe21} This approach exploits
that microstates are typically structurally homogeneous, and therefore
require only a relatively short lag time to randomize and show Markovian behavior. Macrostates, on the other hand, typically contain numerous
structurally different microstates that may be separated by free
energy barriers, resulting in long lag times to reach Markovianity. Along these lines, Hummer and Szabo \cite{Hummer15} proposed a valuable
formulation to optimally project the microstate dynamics onto the
macrostate dynamics, which was shown to achieve significantly improved
Markovianity.

Alternatively, we may consider an (in principle) exact equation of
motion for the time-dependent state vector $\vec{P}(t)$.
\cite{Zwanzig83,Cao20,Hartich21,Suarez21,Dominic23} By employing
projection operator techniques, Zwanzig \cite{Zwanzig83} derived the
generalized master equation
\begin{equation} \label{eq:GME}
\frac{d \vec{P}(t)}{dt} = \int_0^t \vec{K}(t') \vec{P}(t - t') \, dt' ,
\end{equation}
where $\vec{K}(t)$ is the memory kernel matrix that accounts for the
non-Markovian dynamics of the system. To obtain an equation of motion
for the transition matrix, we may replace $\vec{P}(t)$ by $\vec{T}(t)$
using Eq.\ (\ref{eq:Pt}). While the estimation of the memory kernel
for (typically noisy and under-sampled) MD data represents a
non-trivial task,\cite{Jung17,Kowalik19,Meyer20} Huang and
coworkers\cite{Cao20} recently proposed a promising method termed
quasi-MSM (qMSM), which was successfully applied to various
systems.\cite{Unarta21,Cao23}
We also wish to mention a simpler but often effective way of taking
memory into account. By `coring' the macrostate trajectory, we either
request that a transition from one state to another must reach some
pre-defined core region of the other
state,\cite{Buchete08,Schuette11,Lemke16} or that the system spends a
certain minimum time in the new state \cite{Jain14}. The latter
approach, termed `dynamical coring' was shown to considerably improve
the Markovianity of the resulting metastable states.\cite{Nagel19}

In this work, we discuss the theoretical basis and the performance of
various approaches to estimate $\vec{T}(t)$, including the Hummer-Szabo
projection,\cite{Hummer15} the qMSM ansatz of Huang and
coworkers,\cite{Cao20} and two new approaches termed
`microstate-based' and `hybrid MD/MSM' method, respectively. Adopting
a one-dimensional toy model and an all-atom folding trajectory of
HP35,\cite{Piana12} we discuss the applicability as well as the
virtues and shortcomings of these approaches.

%
%
\section{Theory and Methods}

To estimate the macrostate transition matrix $\vec{T}$, we consider
three complementary approaches, which use ({\bf A}) the dynamics of the
microstates, ({\bf B}) the qMSM approximation of the generalized master
equation, and ({\bf C}) a hybrid MD/MSM formulation.

\subsection{From micro- to macrostate dynamics}

In a typical MSM workflow we first construct (say, $n \!\sim\! 10^2$ -
$10^3$) microstates from the MD data, which are subsequently lumped
into a few (say, $N \lesssim 10$) macrostates. As explained in the
Introduction, we assume that the microstates show Markovian behavior at the
chosen lag time $\tau$, which is sufficiently short to
resolve the fastest dynamics of interest. Hence, the microstate
transition matrix $\vec{t} = \{t_{ij}\}$ satisfies the
Chapman-Kolmogorov relation\cite{note2}
\begin{equation} \label{eq:microCK}
\vec{t}(m\tau) = \vec{t}^m (\tau)\, ,
\end{equation} 
that is, the microstate MSM with lag time $\tau$ reproduces
correctly the time evolution of the microstate population vector
$\vec{p}(t) = (p_1,\ldots,p_n)^T$. Here the microstate transition
matrix $\vec{t}(\tau)$ is estimated from the MD data,
simply by counting the transitions from state $j$ to state $i$ within
lag time $\tau$.

We now combine the above microstates into macrostates $J=1,\ldots,N$
($N \!\ll\! n$), such that each macrostate $J$ contains a specific set
of microstates $\{j\}_J$. The macrostate populations $P_J(t)$
are therefore given by
\begin{equation}\label{eq:macroPops}
 P_J(t) = \sum_{j \in J} p_j(t)  \,.
\end{equation}
As for the microstates, we may calculate the corresponding macrostate
transition matrix elements $T_{IJ}(\tau)$ from the MD macrostate trajectory by
counting the transitions from state $J$ to state $I$ within
time $\tau$. Contrary to the microstate MSM, however, the resulting
Chapman-Kolmogorov relation for the macrostates
is generally not valid, because no clear time separation between
intra- and inter-state transitions exists. Hence this
`local-equilibrium' approximation of the macrostate transition
matrix, $\vec{T}_{\rm LE}(\tau)$, cannot correctly reproduce the
macrostate populations $P_J(t)$ as found in the MD simulations.

To improve the calculation of the macrostate transition matrix, we follow
Hummer and Szabo \cite{Hummer15} and introduce the aggregation matrix
$\vec{A} \in \mathbb{R}^{n\times N}$ with elements $A_{iJ}=1$ (if
$i \in J$) and $A_{iJ}=0$ otherwise. This allows us to rewrite Eq.\
(\ref{eq:macroPops}) in vectorial notation,
\begin{equation}\label{eq:macroPops2}
 \vec{P}(t) =  \vec{A}^T\vec{p}(t)  \,.
\end{equation}
To derive a similar relation between the $n\!\times \!n$ matrix
$\vec{t}$ and the $N\!\times \!N$ matrix $\vec{T}$, we define the
diagonal matrices $\vec{D}_n \!\in\! \mathbb{R}^{n\times n}$ and
$\vec{D}_N \!\in\! \mathbb{R}^{N\times N}$, whose elements are given
by the normalized equilibrium populations $p_i^{\rm eq} \equiv \pi_i $
and $P_J^{\rm eq} \equiv \Pi_J$, respectively.\cite{Hummer15} Using
$\vec{D}_N$ and $\vec{D}_n$ to normalize the aggregation matrices
$\vec{A}$ and $\vec{A}^T$, we find
$\vec{A}^T \vec{D}_n \vec{A} \vec{D}_N^{-1} = \vec{1}_N$, as can be
verified by insertion. This leads to the desired relation between
micro- and macrostate transition matrices 
\begin{equation}\label{eq:macroTPM3}
\vec{T}(t)  = \vec{A}^T  \vec{t}(t)\vec{D}_n  \vec{A} \vec{D}_N^{-1} \, .
\end{equation}

It can be evaluated in various ways. In the local-equilibrium
approximation, we first calculate the microstate transition matrix
$\vec{t}(\tau)$, perform the
transformation to get the corresponding macrostate transition
matrix $\vec{T}(\tau)$, and arrive at the MSM expression
\begin{equation}\label{eq:macroTPM4}
\vec{T}_{\rm LE}(m \tau)  = \vec{T}_{\rm
  LE}^m(\tau)  = \left[\vec{A}^T 
  \vec{t}(\tau) \vec{D}_n \vec{A} \vec{D}_N^{-1}\right]^m  \,.
\end{equation}
That is, to evaluate $\vec{T}_{\rm LE}^m(\tau)$, in effect we only
need the definition of the macrostates.

What is more, Eq.\ (\ref{eq:macroTPM3}) suggest also an alternative and
potentially better approximation of $\vec{T}(t)$.
Since we initially assumed that microstate transition matrix
$\vec{t} (\tau)$ satisfies the Chapman-Kolmogorov relation
in Eq.\ (\ref{eq:microCK}), we may first calculate the microstate time
evolution via Eq.\ (\ref{eq:microCK}) and subsequently perform the
transformation to the macrostate transition matrix, i.e.
\begin{equation}\label{eq:DmacroTPM}
\vec{T}_{\rm Mic}(m \tau)  = \vec{A}^T
\vec{t}^m(\tau) \vec{D}_n \vec{A} \vec{D}_N^{-1} \, .
\end{equation}
This microstate-based evaluation of the macrostate transition matrix
preserves all dynamical properties established for the microstates.
For example, it yields by design exact macrostate population dynamics,
and also provides correct state-to-state waiting times and transition
pathways from a microstate Monte Carlo Markov chain simulations with
subsequent projection on the macrostates. While the definition of
$\vec{T}_{\rm Mic}$ is straightforward and its virtues are clearly
promising, this microstate-based approach has to the best of our
knowledge not yet been mentioned.
Compared to the local-equilibrium approximation that yields a constant
transition matrix $\vec{T}_{\rm LE}(\tau)$ and the MSM relation
(\ref{eq:macroTPM4}), however, the microstate-based evaluation results
in a time-dependent macrostate transition matrix.

Alternatively, Hummer and Szabo \cite{Hummer15} used a Laplace
transformation to derive a long-time approximation of the
macrostate transition matrix, which leads to an optimal projection
of the microstate dynamics onto the macrostate dynamics. They obtained
$\vec{T}_{\rm HS}(m \tau) = \vec{T}_{\rm
  HS}^m(\tau)$ with 
\begin{equation}  \label{eq:HS}
 \vec{T}_{\rm HS}(\tau) =  \vec{1}_N + \vec{D}_N - \vec{D}_N \!
\left[ \vec{A}^T \!\left(\vec{1}_n \!+\! \vec{D}_n \!- \!
\vec{t}(\tau)\right)^{-1} \vec{D}_n \vec{A} \right]^{-1} ,
\end{equation}
which provides a constant transition matrix
$\vec{T}_{\rm HS}(\tau)$, but requires the inversion of
$n$- and $N$-dimensional matrices.

Hence, we have introduced three ways to calculate
$\vec{T}(m \tau)$ from $\vec{t}(\tau)$: The
standard local-equilibrium approximation [Eq.\ (\ref{eq:macroTPM4})],
the Hummer-Szabo expression [Eq.\ (\ref{eq:HS})], and the microstate-based
evaluation [Eq.\ (\ref{eq:DmacroTPM})].

%
%
\subsection{qMSM evaluation of the generalized
  master equation}

Huang and coworkers\cite{Cao20} recently derived a generalized
master equation for the transition matrix $\vec{T}(t)$ of the form
\begin{equation} \label{eq:GME_Huang}
    \Dot{\vec{T}}(t) = \Dot{\vec{T}}(0)\vec{T}(t) + \int_0^{\min [t,
      \tau_{\mathrm{K}}]} \!\!\!\!\!\!\!\!\!\!\!\!\!\!\!\!\!\!
    \vec{K}(t') \, \vec{T}(t\!-\!t') dt'. 
\end{equation}
Compared to Eq.\ (\ref{eq:GME}), it contains the additional term
$\Dot{\vec{T}}(0)\vec{T}(t)$, which reflects the time evolution of the
system without interaction with the environment. Moreover, the upper
bound of the integral assumes that the memory kernel $\vec{K}(t)$
decays within the (presumably short) time $\tau_{\mathrm{K}}$.

To solve Eq.\ (\ref{eq:GME_Huang}), we discretize the equation by
introducing $t \!=\! t_n \!=\! n \Delta t$, 
$\vec{T}(t_n) \!=\! \vec{T}_n$ and $\vec{K}(t_n) \!=\! \vec{K}_n$, yielding
\begin{equation} \label{eq:qmsm_T}
  \Dot{\vec{T}}_n = \Dot{\vec{T}}_0 \vec{T}_n +
  \Delta t \sum_{m=1}^{n} \vec{K}_n \vec{T}_{n-m}.    
\end{equation}
The main idea of the qMSM approach is to first obtain a short-time
approximation of the transition matrix from the MD data,
$\vec{T} \approx \vec{T}_{\rm MD}$. Subsequently,
$\vec{T}_{\rm MD}$ is used to iteratively solve Eq.\
(\ref{eq:GME_Huang}) for the memory matrix via  
\begin{equation} \label{eq:qmsm_K}
 \vec{K}_n = \frac{\Dot{\vec{T}}_n  
            - \Dot{\vec{T}}_0 \vec{T}_n} {\Delta t}
          - \sum_{m=1}^{n-1} \vec{K}_m \,\vec{T}_{n-m} \, .
\end{equation}
Inserting the resulting $\vec{K}_n$ in Eq.\ (\ref{eq:qmsm_T}), we
obtain the desired evolution of $\vec{T}(t)$ for long times
($t \gg \tau_{\mathrm{K}}$). Hence the qMSM formulation facilitates a
straightforward approximate solution of an in principle exact
generalized master equation, and also explicitly provides the memory
matrix $\vec{K}(t)$ that accounts for the non-Markovian dynamics of
the system. Requiring only MD data of length
$t \approx \tau_{\mathrm{K}}$, qMSM allows us to predict long-time
dynamics from short trajectories.
The above equations were implemented in an in-house Python
code following Ref.\ \onlinecite{Wu24}; the kernel decay time
$\tau_{\mathrm{K}}$ was obtained either directly from the memory kernel
  (in the one-dimensional toy model), or via the mean integral memory
  kernel\cite{Cao20} (for HP35).

While the qMSM inversion scheme [i.e., solving Eq.\
(\ref{eq:qmsm_K}) to get $\vec{K}$ and inserting it in Eq.\
(\ref{eq:qmsm_T}) to get $\vec{T}$] appears to be a straightforward
data-driven approach to parameterize the generalized master equation,
there are several peculiarities of the qMSM method.
For one, we note that --given long enough MD data-- the MD-based
matrix $\vec{T}_{\rm MD}$ is the best transition matrix you can
get. Hence, by calculating $\vec{K}$ from $\vec{T}_{\rm MD}$ via Eq.\
(\ref{eq:qmsm_K}) and inserting it in Eq.\ (\ref{eq:qmsm_T}), the
resulting transition matrix $\vec{T}$ can only be worse than the
initial matrix $\vec{T}_{\rm MD}$. Iterating the procedure [by using
the new $\vec{T}$ to again calculate $\vec{K}$ and then again
$\vec{T}$] does not help either, because we obtain the same result as
in the first iteration (as readily shown by insertion).


Another issue is that qMSM keeps the additional term
$\Dot{\vec{T}}(0)\vec{T}(t)$, which typically vanishes for symmetry
reasons,\cite{Zwanzig83} but may be non-zero in a numerical
evaluation. Interestingly, this term exactly cancels the memory kernel
at $t\!=\!0$ [which can be shown by calculating $\vec{T}_n$ and
$\vec{K}_n$ for $n=0$ from Eqs.\ (\ref{eq:qmsm_T}) and
(\ref{eq:qmsm_K}), respectively]. While this is again a consequence of
the chosen discretization, it seems odd, because $\vec{K}_0$ is the
only non-zero term in the Markov limit. In all cases considered,
however, the neglect of the term seems to hardly change the numerical
results.
Another practical issue is that the calculation of the memory matrix
from Eq.\ (\ref{eq:qmsm_K}) can be seriously plagued by noise resulting
from the MD data and the time derivatives. As a remedy, recently 
an integrative ansatz \cite{Cao23} and a time-convolutionless approach
\cite{Dominic23} of the generalized master equation were proposed.

%
\subsection{Hybrid MD/MSM formulation}

In the Introduction, we opposed in Eqs.\ (\ref{eq:Pt}) and
(\ref{eq:MSM}) two standard ways to estimate the macrostate transition
matrix $\vec{T}(t) = \{T_{IJ}(t)\}$. On the one hand, we may use the
MD data to directly count the transitions from state $J$ to state $I$ within
time $t$. The resulting transition matrix $\vec{T}_{\rm MD}(t)$
generates the correct time evolution of the state model according to
Eq.\ (\ref{eq:Pt}), but is limited to times $t \approx t_{\rm max}$
with $t_{\rm max}$ being the length of the MD trajectories (and
actually even some hundreds MD frames less to facilitate a sufficient
time average). 
On the other hand, we may invoke the Markov approximation to obtain
$\vec{T}(t\!=\!m\tau) \approx \vec{T}^m(\tau)$, which assumes that the
transition matrix $\vec{T}(t)$ becomes constant for $t \gtrsim \tau$.
While this only requires short MD trajectories, we need to choose a lag
time $\tau$ that is long enough to achieve Markovianity, but at the
same time short enough to resolve the fastest dynamics of interest.

To have the best of both worlds, we may simply combine the two
approaches by using $\vec{T}_{\rm MD}(t)$ at short times and employing
some formulation (including local equilibrium, Hummer-Szabo, or
microstate-based) to construct a MSM using the long lag time
$t_{\rm max}$, i.e.,
\begin{equation} \label{eq:hybrid}
  \vec{T}_{\rm MD/MSM}(t) =
    \begin{cases}
      \vec{T}_{\rm MD}(t) & t \le \,t_{\rm max},\\
      \vec{T}_{\rm MSM}^m(t_{\rm max}) & t\!=m t_{\rm max} .
    \end{cases}       
\end{equation}
In this way, we use at short times the approximation-free transition
matrix $\vec{T}_{\rm MD}(t)$, including full time resolution (as given
by the MD data). At long times, the MSM typically can use a rather
long lag time.

%
%
\section{Results}

To compare the performance of the different methods, we test them on
two model systems: A one-dimensional toy model lumping four
microstates into two macrostates, and a recently
established\cite{Nagel23b} benchmark MSM of the folding of villin
headpiece (aka HP35) using a long MD trajectory by Piana et
al.\cite{Piana12}

\subsection{One-dimensional toy model}
\label{sec:toymod}

\begin{figure*}[t!]
    \centering
    \includegraphics[width=0.9\textwidth]{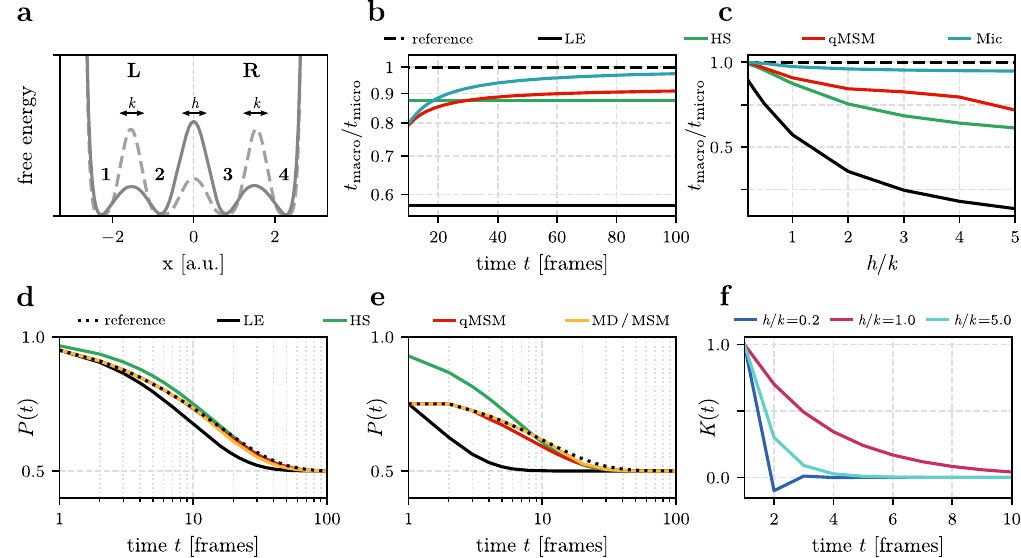}
    \caption{
      One-dimensional toy model. (a) Schematic free energy landscape,
      indicating four microstates 1, 2, 3 and 4, which are lumped into
      two macrostates $L$ and $R$. Depending on the ratio of the
      interstate transition probability \textit{h} and the intrastate
      transition probability \textit{k}, the macrostate dynamics is
      Markovian ($h/k \ll 1$, dark gray) or non-Markovian
      ($h/k \gtrsim 1$, dashed). (b) Ratio of the macroscopic and
      microscopic implied timescales $t_{\rm macro}/t_{\rm micro}$ of
      the two-macrostate model for $h/k=1$. Compared are the reference
      result obtained from the microstates (dashed black), the local
      equilibrium approximation (black), and the Hummer-Szabo
      projection (green), as well as the time evolution the result
      from qMSM (red) and of
      the microstate-based result (blue). (c) Normalized implied
      timescales obtained from the various methods, drawn as a
      function of the Markovianity parameter $h/k$. Time evolution of
      the macrostate population of the various methods, shown for (d)
      $h/k=1$ and (e) $h/k=5$. Also shown are results of the hybrid
      MD/MSM calculation [Eq.\ (\ref{eq:hybrid})] with
      $t_{\rm max}=10$. (f) Normalized memory kernel $K(t)$ of the
      qMSM method, obtained for $h/k=0.2$ (blue), $h/k=1$ (red), and
      $h/k=5$ (green).}
    \label{fig:1Dmodel}
\end{figure*}

As a simple instructive model, Fig.\ \ref{fig:1Dmodel}a shows the free
energy landscape of a system with four 
microstates $i$ ($i=1,\ldots,4$), which are lumped into two
macrostates $I$ ($I=L,R$ for left and right state). Following
Hummer and Szabo, \cite{Hummer15} we define the model via its
microstate transitions
\[ 1 \xleftrightarrow{k} 2 \xleftrightarrow{h} 3 \xleftrightarrow{k} 4
\]
with transition probabilities \textit{k} and \textit{h}. 
The resulting microstate transition matrix reads
\begin{equation}
\vec{t} = 
\begin{bmatrix}
1 \!-\! k & k & 0 & 0 \\
k & 1 \!-\! (k \!+\! h) & h & 0 \\
0 & h & 1 \!-\! (h \!+\! k) & k \\
0 & 0 & k & 1 \!-\! k
\end{bmatrix} .
\end{equation}
Because $\vec{t}$ is symmetric and detailed balance holds
($t_{ij} \pi_j \!=\! t_{ji} \pi_i$), the microstate equilibrium
populations are $\pi_i \!=\! 1/4$. The eigenvalues of the matrix are
$\lambda_0=1, \lambda_1=1\!-\!h\!-\!k\!+\! \sqrt{h^2\!+\!k^2},
\lambda_2=1\!-\!2k$, and
$\lambda_3=1\!-\!h\!-\!k\!-\! \sqrt{h^2\!+\!k^2}$. Hence, the longest
(nontrivial) microstate implied timescale is
\begin{equation}
t_{\rm micro} \equiv t_1/\tau_0 = - \frac{1}{\ln \lambda_1} = - \frac{1}{\ln (1\!-h\!-k\!+\!
\sqrt{h^2\!+\!k^2})},
\end{equation}
where we introduced the unit time $\tau_0$ to define a dimensionless
time $t_{\rm micro}$.

When we combine microstates 1 and 2 into macrostate $L$ and
microstates 3 and 4 into macrostate $R$, the probability \textit{k}
accounts for intrastate transitions (i.e., within $L$ or $R$) and
\textit{h} accounts for interstate transitions (i.e., between $L$ and
$R$) of the macrostate model, see Fig.\ \ref{fig:1Dmodel}a. We study
the model both in the case of good lumping (i.e., a sufficient
timescales separation between fast intrastate and slow interstate
motions) and in the case of bad lumping. The latter means that we have
a low energy barrier between macrostates $L$ and $R$, and high
barriers between microstates that belong to the same macrostate. This
leads to a bad timescales separation, causing the model to behave
non-Markovian.

Due to the simplicity of the model, we can derive analytical
expressions for the $2 \!\times\! 2$ macrostate transition matrix. In
the local equilibrium approximation [Eq.\ (\ref{eq:macroTPM4})], we
obtain
\begin{equation}
\vec{T}_{\text{LE}}(t\!=\!m \tau_0) = 
\begin{bmatrix}
1 - h/2 &  h/2\\
h/2 & 1 - h/2
\end{bmatrix}^m .
\end{equation}
Notably, $\vec{T}_{\text{LE}}$ does not depend on the intrastate
transition rate $k$, as it is assumed to be much higher than any other
timescale of the macrostate model. The single (nontrivial) macrostate
implied timescale is 
\begin{equation}
t_{\rm LE} = - \frac{1}{\ln (1-h)}.
\end{equation}
The Hummer-Szabo projection
[Eq.\ (\ref{eq:HS})] gives $\vec{T}_{\rm HS}(t\!=\!m \tau_0) = \vec{T}_{\rm
  HS}^m$ with \cite{Hummer15}
\begin{equation}
\vec{T}_{\text{HS}} = \vec{1}_2 + \frac{h k}{h+2k}
\begin{bmatrix}
-1 & 1 \\
1 & -1
\end{bmatrix},
\end{equation}
which takes into account both intra- and inter-state transitions.
The resulting implied timescale is 
\begin{equation}
t_{\rm HS} = - \frac{1}{\ln \left(1-\frac{2h k}{h+2k} \right)}.
\end{equation}
In the microstate-based approach [Eq.\ (\ref{eq:DmacroTPM})] and the
qMSM method [Eq.\ (\ref{eq:qmsm_T})], the transition matrices
$\vec{T}_{\text{Mic}}(t)$ and $\vec{T}_{\text{qMSM}}(t)$ are
calculated numerically, because no simple analytical expressions
exist. Note that the implied timescales of $\vec{T}_{\text{Mic}}(t)$
and $\vec{T}_{\text{qMSM}}(t)$ depend on time, while the implied
timescales $t_{\rm micro}$, $t_{\rm LE}$, and $t_{\rm HS}$ are constant.

As a first test, we wish to study how the various formulations
reproduce the reference implied timescale $t_{\rm micro}$ of the
microstates. Choosing $k=h=0.1$ (weakly non-Markovian case), Fig.\
\ref{fig:1Dmodel}b shows that the local equilibrium approximation
$t_{\rm LE}$ underestimates $t_{\rm micro}$ almost by a factor 2,
while the Hummer-Szabo result is only shorter by 13\,\%.  The results
obtained for $\vec{T}_{\text{Mic}}(t)$ and $\vec{T}_{\text{qMSM}}(t)$
reveal that the microstate-based method converges to the correct
result and that qMSM converges to a value that is 9\,\% shorter than
$t_{\rm micro}$. (In all cases, we choose the qMSM kernel decay time
$\tau_{\rm K}$ such that the normalized memory kernel satisfies
$K(\tau_{\rm K}) \le 0.1$.)
To compare to a largely Markovian case ($h/k=0.2$), Fig.\
\SIfigITScontrol{} shows that all methods yield the correct
timescale $t_{\rm micro}$, except for the local equilibrium
approximation, which gives $t_{\rm LE} = 0.9\, t_{\rm micro}$.

We now choose $t=100$ (where qMSM and microstate-based results are
virtually constant), and vary the Markovianity parameter $h/k$ from
$h/k \ll 1$ (Markovian case) to $h/k \gtrsim 1$ (non-Markovian
case). While for $h=0.1,\, k=0.5$ all methods yield the correct microstate
timescale $t_{\rm micro}$ (except for $t_{\rm LE}$), the results vary
considerably in the non-Markovian case. Choosing $h=0.5,\, k=0.1$, for example,
$t_{\rm LE}$ underestimates $t_{\rm micro}$ by 87\,\%, $t_{\rm HS}$ by
40\,\%, and $t_{\rm qMSM}$ by 27\,\%, while the microstate-based result
closely matches the reference.

We now consider the quality of the underlying time-dependent
transition matrices. As the diagonal elements $T_{II}(t)$ ($I=L,R$)
represent the state populations $P_L(t) = 1\!-\!P_R(t)$, which also
determine the off-diagonal elements $T_{LR}=1\!-\!P_L$ and
$T_{RL}=1\!-\!P_R$, it is sufficient to focus on $P_L(t) \equiv
P(t)$. By design, $P(0)=1$ and $P(\infty)=\Pi_L=0.5$, such that the
various formulations only differ by the rate of the decay. Because the
microstates produce the exact dynamics of the model, the microstate-based
calculation $P_{\rm Mic}(t)$ coincides with this reference.

Choosing $h/k=1$ and $h/k=5$, Figs.\ \ref{fig:1Dmodel}d and
\ref{fig:1Dmodel}e compare the population $P(t)$ obtained from the
various formulations. While for $h/k=1$ all methods are
quite close to the reference results, we find significant deviations
in the strongly non-Markovian case $h/k=5$, where $P(t)$ in fact
reveals two timescales. That is, the population exhibits a strong
initial decay to $P(1)=0.75$ during the first time step, reflecting
fast transitions over the shallow barrier at $x=0$ (Fig.\
\ref{fig:1Dmodel}a). All formulations reproduce this value correctly,
except for the long-time approximation $P_{\rm HS}(t)$, which
coincides with the reference for $t \gtrsim 7$. Being a short-time
approximation, $P_{\rm LE}(t)$ decays significantly too fast for
$t>1$. The qMSM calculation represents a clear improvement over these
approximations and reproduces the exact results closely. Finally, we
also show the hybrid calculation [Eq.\ (\ref{eq:hybrid})], which uses
until $t_{\rm max}=10$ the exact transition matrix
$\vec{T}_{\rm MD}(t)$, and for longer times the local approximation
$\vec{T}_{\rm LE}(t = m t_{\rm max})$. For this choice of
$t_{\rm max}$, the hybrid result virtually matches the reference at
all times.

Unlike the other methods, the generalized master equation
(\ref{eq:GME_Huang}) underlying the qMSM method formulation involves the calculation of a
memory matrix $\vec{K}(t)$, which reports on the non-Markovianity of
the dynamics.  Since for symmetry reasons
$K_{LL}=K_{RR}=-K_{LR}=-K_{RL}$, it is sufficient to focus on
$K_{LL}(t)$. Considering the cases $h/k=0.2,\, 1$, and 5, Fig.\
\ref{fig:1Dmodel}f shows the normalized memory kernel
$K(t)= K_{LL}(t)/K_{LL}(1)$. As expected, in the Markovian limit
($h/k=0.2$) the memory decays essentially within a single time
step. Less expected, though, we find that the slowest decay
($\sim 2.8$) is obtained for weakly non-Markovian case ($h/k=1$), while
in the strongly non-Markovian case ($h/k=5$), the decay ($\sim 0.8$) is
again faster, although clearly slower than the Markovian case.

To explain this finding, we recall that the memory kernel $K(t)$ is
meant to reflect the intrastate dynamics of the macrostates, which in turn
depends on the implied timescales $t_k$ ($k=1,\,2,\,3$) and associated
eigenvectors of the microstate MSM. As shown in Tab.\ \SItabmemory, the first
timescale and eigenvector clearly indicate transitions between the
left states ($L=1,\,2$) and the right states ($R=3,\,4$) , while the second
and third eigenvectors reflect transitions within $L$ and $R$.
In the Markovian limit ($h/k=0.2$), we find $t_2=0.45$ and $t_3=0.26$,
which indeed coincide with the decay of the memory kernel $K(t)$ in
Fig.\ \ref{fig:1Dmodel}d. Moreover, for $h/k=1$, we obtain $t_2=4.5$
and $t_3=2.4$, which again are similar to the decay time ($\sim 2.8$)
of $K(t)$. In the strongly non-Markovian case ($h/k=5$), however, this
coincidence is less clear, as we get $t_2=4.5$ and $t_3=0.45$, but a
memory decay time of 0.8. Interestingly, we also find that for $h/k=5$
the amplitude $K(1)$ of the memory is about twenty times larger than
in the first two cases. These findings indicate that the
interpretation of the qMSM memory kernel in terms of memory timescales
is generally not straightforward.\cite{Vollmar24}

%
%
\subsection{The folding of HP35}

\begin{figure*}[t!]
    \includegraphics[width=0.9\textwidth]{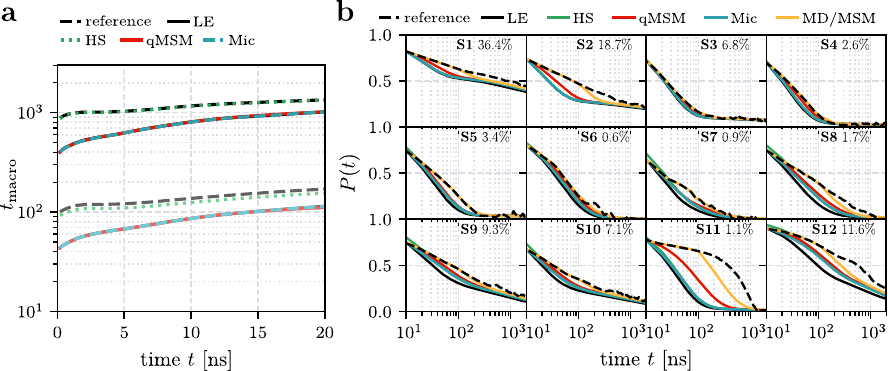}
    \caption{
      Twelve-state MSM constructed from a $300 \mu$s-long MD
      trajectory\cite{Piana12} of the folding of HP35. (a) First two
      implied timescales obtained from the microstates (dashed black,
      reference), the local equilibrium approximation (using the lag
      time $\tau=10\,$ns, black), the Hummer-Szabo projection (using
      $\tau=10\,$ns, green), the microstate-based result (blue), and
      the result from qMSM (using the kernel time
      $\tau_{\rm K}=10\,$ns, red). (b) Chapman-Kolmogorov test of the
      twelve macrostates, comparing the reference MD state trajectory
      to the various versions of the theory. Also shown are results of
      the hybrid MD/MSM calculation [Eq.\ (\ref{eq:hybrid})] with
      $t_{\rm max}=100\,$ns.}
    \label{fig:HP35}
\end{figure*}

To learn how the above findings for the toy model generalize to the
case of all-atom MD data, we now consider the folding of villin
headpiece (HP35) as a well-established model problem. As in previous
work,\cite{Nagel23,Nagel23b} we use a $\sim 300 \mu$s-long MD trajectory of the
fast folding Lys24Nle/Lys29Nle mutant of HP35 at $T = 360\,$K by Piana
et al., \cite{Piana12} which shows about 30 folding events.
Following the benchmark study of Nagel et al.,\cite{Nagel23b} we employ
the following MSM workflow. Firstly, we use
MoSAIC correlation analysis\cite{Diez22} to perform a feature
selection resulting in 42 contact distances. Next, we eliminate
high-frequency fluctuations of the distance trajectory, by employing a
Gaussian low-pass filter\cite{Nagel23} with a standard deviation of
$\sigma=2\,$ns. Employing principal component analysis on these
contacts,\cite{Ernst15} we use the resulting first five components for
subsequent robust density based clustering\cite{Sittel16} into 547
microstates. In a final step, we adopt the most probable path
algorithm\cite{Jain12} to dynamically lump the microstates into
12 metastable macrostates, using a lag time of
10\,ns. The resulting dynamical model
compares favorably to MSMs from alternative combinations of
methods\cite{Nagel23b} and explains the folding of HP35 as cooperative
transition between folded and unfolded energy basins and several
intermediate states. In this work, we employ the partitioning of the MD
trajectory into micro- and macrostates, in order to calculate the
various versions of the macrostate transition matrix introduced above.

\begin{figure*}[t!]
	\centering
	\includegraphics[width=0.9\textwidth]{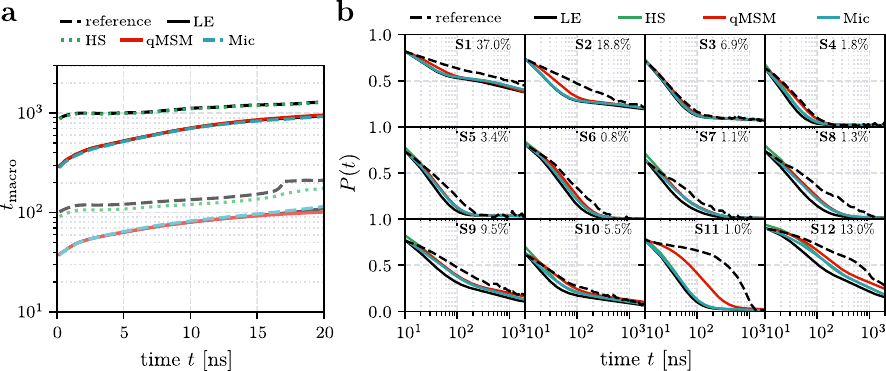}
	\caption{As in Fig.\ \ref{fig:HP35}, except that the original
		$300 \mu$s-long trajectory of HP35 was split in
		$10~000 \times 30\,$ns trajectories. The hybrid MD/MSM method is
		not shown, as for $t_{\textit{max}} = 10\,$ns it coincides with
		the local equilibrium approximation.}
	\label{fig:HP35_split}
\end{figure*}

We begin with the discussion of the first two implied timescales,
which are depicted in Fig.\ \ref{fig:HP35}a as a function of the lag
time. As discussed in Ref.\ \onlinecite{Nagel23b}, the slowest
timescale ($\sim 1.5\,\mu$s) corresponds to the overall folding
process, while the next two timescales ($\sim 0.1\,\mu$s) account for
conformational rearrangement in the native and the unfolded basins.
Taking again the microstate timescales as a reference, we overall find
the Hummer-Szabo results being in excellent agreement, followed by the
results from the local equilibrium approximation, which virtually
coincide with the qMSM and the microstate-based results.

As most important assessment, Fig.\ \ref{fig:HP35}b shows the
Chapman-Kolmogorov test of the twelve macrostates, which compare the
macrostate projection of the MD trajectory (providing the reference)
to the various versions of the theory. As a result of a careful
selection of features and methods, the benchmark model\cite{Nagel23b}
represents a valuable MSM. Consequently, the deviations of the
various methods are overall only minor, and already the simple local
equilibrium approximation matches the correct population dynamics
of most macrostates quite closely. Quite similar is the
Hummer-Szabo projection, despite its superior modeling of the implied
timescales above.

Recalling that the microstate-based calculation should coincide with the
reference MD results in the case of a truly Markovian partitioning of
the microstates, we learn that the microstates are a main reason for
the deviations of the Chapman-Kolmogorov tests. Not relying on the
quality of the microstates, the qMSM in fact represents an improvement
of the microstate-based methods, particularly in the case of
ill-defined states such as states 2 and 11. (Judging from the time
evolution of the qMSM memory matrix $\vec{K}(t)$ shown in Fig.\
\SIfigVillinMemory, we chose $\tau_{\rm K}=10\,$ns where, apart from
residual fluctuations around zero, all elements
$K_{IJ}(t)$ are safely decayed.)
Even better is the performance of a hybrid calculation [Eq.\
(\ref{eq:hybrid})], which uses until $t_{\rm max}=100\,$ns the exact
transition matrix $\vec{T}_{\rm MD}(t)$, and for longer times the local
approximation $\vec{T}_{\rm LE}(t = m t_{\rm max})$. Apart from the 
ill-defined states 2 and 11, the hybrid calculation matches the
reference very accurately.

While a single, long MD trajectory is certainly perfect for building an
MSM, in practice we often have many short trajectories (since they
are readily computed in parallel). In principle, trajectories as short
as the lag time $\tau$ and the kernel time $\tau_{\rm K}$ should be
already sufficient to construct an MSM and a qMSM,
respectively. However, the validity of this presumption has been
rarely assessed due to the lack of long reference trajectories. As we
used $\tau=\tau_{\rm K}=10\,$ns in the calculations above, and we need
some extra time for the time averages to calculate transition
probabilities, we decided to split the $300 \mu$s trajectory in
$10000 \times 30\,$ns (150 frames) long pieces, such that we can
average over 100 frames. Note that 30\,ns is quite short considering
the overall folding time of $\sim 2\,\mu$s of HP35. As the MSM
workflow explained above employs no time information up to the density
based clustering included, \cite{note1} we can use the same microstates. On the
other hand, the dynamical lumping of these states via the most
probable path algorithm\cite{Jain12} requires calculations based on the
transition matrix, which differs in the case of one long and of many
short trajectories. Hence, the resulting macrostates differ minor from
the ones obtained from the single $300 \mu$s trajectory, see Fig.\
\SIfigShortTrajmodel. Lastly, we again calculated the macrostate
transition matrices for the various theoretical formulations.

Figure \ref{fig:HP35_split} shows the implied timescales and
Chapman-Kolmogorov tests of the resulting MSMs. Overall the
short-trajectories results look very similar to the results from the
single long trajectory (Fig.\ \ref{fig:HP35}), thus confirming the
promise of MSMs to describe long-time dynamics from short
trajectories. However, it should be kept in mind that 
these short trajectories used adequately chosen initial conditions,
because they were taken from a long trajectory that exhibits numerous
folding and unfolding transition. This is less obvious without such a
reactive trajectory, although various strategies have
been proposed to obtain good initial conditions along the reaction
paths of a considered process.\cite{Bowman10a,Biswas18,Wan20}

%
%
\section{Concluding remarks}

We have outlined the theoretical basis of various approaches to
estimate the macrostate transition matrix of an MSM, from the standard
local equilibrium approximation to a sophisticated generalized master
equation approach. Employing a one-dimensional toy model and an
all-atom folding trajectory of HP35, we have provided a comprehensive
comparison of the various formulations, and introduced two new
methods, the microstate-based approach in Eq.\ (\ref{eq:DmacroTPM})
and the hybrid MD/MSM ansatz in Eq.\ (\ref{eq:hybrid}).

The most commonly used method, the local equilibrium approximation
[Eq.\ (\ref{eq:macroTPM4})], simply counts the transitions between the
macrostates within a certain lag time, which assumes Markovian data
with a timescale separation 
between fast intrastate fluctuations and slow interstate transitions.
As shown for the toy model in Fig.\ \ref{fig:1Dmodel}, which allows to
tune the ratio between these timescales, the quality of the resulting implied
timescale and the Chapman-Kolmogorov test readily deteriorate, when this
separation is not fulfilled.

Assuming that Markovianity is at least given at the microstate level,
the knowledge of the microstate dynamics can be exploited to construct
improved macrostate dynamics. For one, this is used in the
Laplace transform-based method by Hummer and Szabo, \cite{Hummer15}
which was shown to achieve significantly improved Markovianity in all
considered cases.
However, we can also directly employ a microstate-to-macrostate
projection to construct a macrostate transition matrix
$\vec{T}_{\rm Mic}(t)$ [Eq.\ (\ref{eq:DmacroTPM})], which by design
yields correct macrostate population dynamics.  As an important
consequence, this means that residual non-Markovian behavior of the
resulting MSM must be caused by suboptimal microstates (and not by the
lumping of micro- into macrostates).  By performing microstate Monte
Carlo Markov chain simulations with subsequent projection on
macrostates, this new and promising microstate-based approach may also
provide state-to-state waiting times and transition pathways.

Not relying on the quality of the microstates, generalized master
equations, such as the qMSM ansatz of Huang and coworkers\cite{Cao20}
provide an alternative approach to deal with non-Markovian dynamics.
In fact, we found that qMSM performs well even in the case of
ill-defined macrostates of HP35 (Fig.\ \ref{fig:HP35}) and in the case
of short-trajectory data (Fig.\ \ref{fig:HP35_split}). We discussed
several peculiarities of the approach, including the interpretation of
the memory kernel matrix and its calculation from noisy data.

In practice, an MSM is often used to interpret given MD data in terms
of the time evolution of the populations $P_I(t)$ of metastable
conformational states. For the length of the MD trajectories
$t_{\rm max}$ (actually some hundreds MD frames less to facilitate a
sufficient time average), we can simply project the MD data on these
states, without invoking a dynamical approximation such as the
assumption of Markovianity. In the `hybrid MD/MSM' method [Eq.\
(\ref{eq:hybrid})], we exploit this idea to obtain the system's short
time evolution with good time resolution, and subsequently use some
formulation (including local equilibrium, Hummer-Szabo, or
microstate-based) to construct a MSM using the long lag time
$t_{\rm max}$. Given sufficiently long trajectories, this strategy of
combining the best of two worlds is simple and promises prime
results. While the definition of the hybrid MD/MSM method and its
virtues seem obvious, it has to the best of our knowledge not yet been
fully appreciated.

%

\subsection*{Acknowledgments}

The authors thank Georg Diez, Daniel Nagel and Tim Uttenweiler for
helpful comments and discussions, as well as D.\ E.\ Shaw Research for
sharing their trajectories of HP35. This work has been supported by
the Deutsche Forschungsgemeinschaft (DFG) within the framework of the
Research Unit FOR 5099 ''Reducing complexity of nonequilibrium''
(project No.~431945604), the High
Performance and Cloud Computing Group at the Zentrum f\"ur
Datenverarbeitung of the University of T\"ubingen and the
Rechenzentrum of the University of Freiburg, the state of
Baden-W\"urttemberg through bwHPC and the DFG through Grant Nos. INST
37/935-1 FUGG (RV bw16I016) and INST 39/963-1 FUGG (RV bw18A004), the Black Forest
Grid Initiative, and the Freiburg Institute for Advanced Studies
(FRIAS) of the Albert-Ludwigs-University Freiburg.

\subsection*{Data Availability Statement}

The simulation data and all intermediate results for our reference
model of HP35, including detailed descriptions to reproduce all steps of the
analyses, can be downloaded from https://github.com/moldyn/HP35.

\vspace*{-4mm}
\begin{suppinfo}
\vspace*{-2mm}
Includes a table comprising the eigenvectors, timescales, and kernel
decay times of all considered one-dimensional models, as well as
figures of the time evolution of the memory kernel of HP35, and of the
structural characterization of the states of the short-trajectory MSM
of HP35.
\end{suppinfo}
\vspace*{-4mm}

%
%
\bibliography{\dir/stock,\dir/md,new}

\begin{mcitethebibliography}{45}
\bibitem[not()]{note2}
{We note that Markovianity is a sufficient condition for validity of the
  Chapman-Kolmogorov equation, not a necessary one. See the discussion in
  Refs.\ \onlinecite{Lapolla21,Vollmar24} }\relax
\mciteBstWouldAddEndPuncttrue
\mciteSetBstMidEndSepPunct{\mcitedefaultmidpunct}
{\mcitedefaultendpunct}{\mcitedefaultseppunct}\relax
\EndOfBibitem
\bibitem[not()]{note1}
{As mentioned, we use a Gaussian low-pass filter ($\sigma=2\,$ns) to smooth the
  input coordinates, which may cause deviations at the beginning and the end of
  the short trajectories, compared to a filtering of the entire trajectory. As
  this effect turned out rather small, though, we ignored this effect and used
  the same microstates for better comparability}\relax
\mciteBstWouldAddEndPuncttrue
\mciteSetBstMidEndSepPunct{\mcitedefaultmidpunct}
{\mcitedefaultendpunct}{\mcitedefaultseppunct}\relax
\EndOfBibitem
\end{mcitethebibliography}

\end{document}


\clearpage
\begin{table}[hb!]
\centering
\renewcommand{\arraystretch}{1.2} 
\setlength{\tabcolsep}{10pt}      

\begin{tabular}{>{\centering\arraybackslash}m{3cm}|
                >{\centering\arraybackslash}m{3cm}
                >{\centering\arraybackslash}m{3cm}
                >{\centering\arraybackslash}m{3cm}} 
\centering & $\mathbf{h=0.1, k=0.5}$ & $\mathbf{k=0.1=h}$ & $\mathbf{h=0.5, k=0.1}$ \\ \hline
\centering $t_1$         & 10.6             & 16.6              & 10.59            \\ 
\centering $t_2$         & 0.45             & 4.48              & 4.48             \\ 
\centering $t_3$         & 0.26             & 2.39              & 0.45             \\ 
\hline
\centering eigenvectors   & \includegraphics[width=0.2\textwidth]{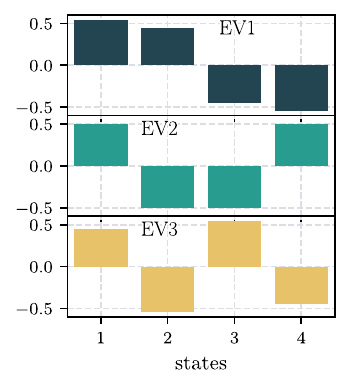} & 
                \includegraphics[width=0.2\textwidth]{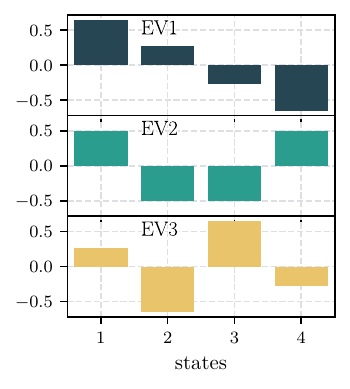} & 
                \includegraphics[width=0.2\textwidth]{TabS1_1.pdf} \\ \hline
\centering $\tau_k$         & 0.45             & 2.8               & 0.83             \\ 
\centering $K(1)$        & 0.005            & 0.005             & 0.12             \\ 
\end{tabular}
\caption{Timescales of the 4-state toy model, obtained for the three cases with different Markovianity. Shown are the three implied timescales $t_\textit{k}$, the corresponding eigenvectors, the decay time of the memory kernel $\tau_k$, and the initial values of the memory kernel $K(1)$.}
\label{tab:timescales_toymodel}
\end{table}
\vspace{1cm}
\begin{figure}[ht!]
	\centering
	{\includegraphics[width=0.7\textwidth]{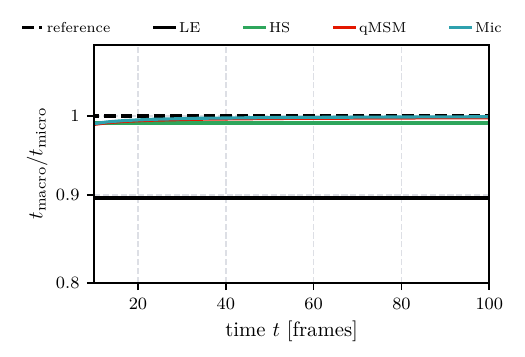}}
	\caption{Implied timescale
		$t_{\rm macro}$ of the two-macrostate toy model for
                the Markovian case $h/k=0.2$. Compared
		are the reference result obtained from the microstates (dashed
		black), the local equilibrium approximation (black), and the
		Hummer-Szabo projection (green), as well as the time evolution
		the result from qMSM (using the Kernel time $\tau_{\rm K}=3$,
			red) and of the microstate-based result (blue).}
	\label{fig1}
\end{figure}

\begin{figure}[ht!]
	\centering
	{\includegraphics[width=0.9\textwidth]{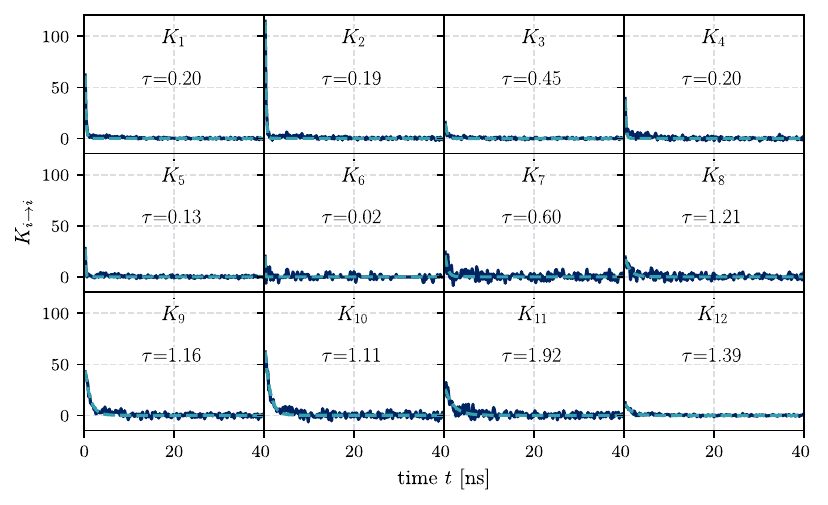}}
	\caption{Diagonal element of the qMSM memory kernel for the 300 $\mu$s trajectory of HP35. The kernels are each fit with an exponential function and the decay times estimated from each fit are shown. The kernels obtained from the 30 ns trajectories are quite similar.}
	\label{fig1}
\end{figure}
\begin{figure}[ht!]
	\centering
	{\includegraphics[width=0.8\textwidth]{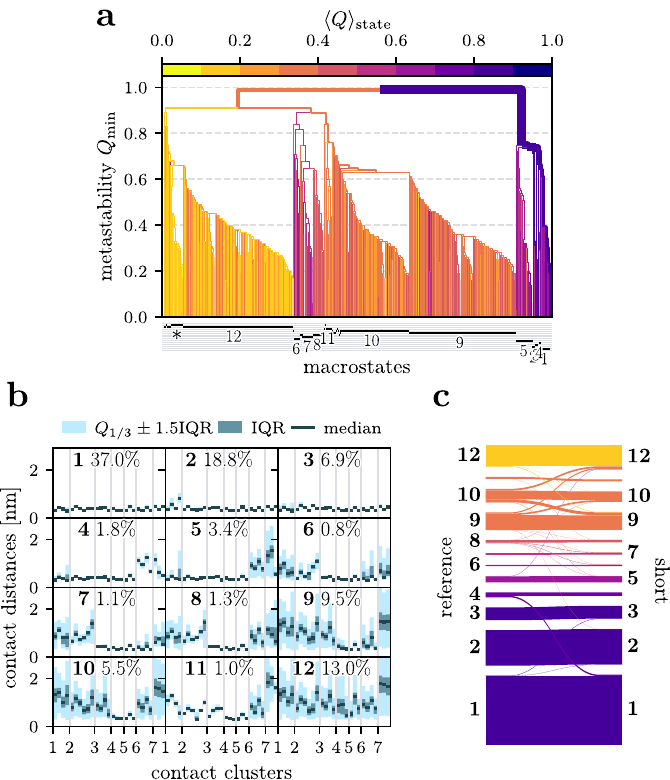}}
	\caption{States analysis for the 12 states obtained by lumping the short microstates trajectories as described in Sect. III.B of the main paper. (a) MPP dendrogram\cite{Jain12} demonstrating the classification of microstates into metastable
states, (b) the contact representation of the resulting metastable states, (c) a Sankey
diagram contrasting the states of the reference model\cite{Nagel23} on the left and the states from
the short trajectories on the right.}
	\label{fig2}
\end{figure}

\clearpage
\bibliography{\dir/stock, \dir/md, \dir/new}